 \documentclass[final,3p,times,sort&compress]{elsarticle}

\usepackage{multirow,setspace,times,amssymb,amsmath,graphicx,color,rotating,subfigure,url}
\usepackage{lineno}
\usepackage{natbib}
\usepackage{booktabs}
\usepackage{hyperref}
\usepackage{threeparttable}

\journal{Physica A}

\begin{document}

\begin{frontmatter}


\title{Impact of information cost and switching of trading strategies in an artificial stock market}
\author[CME,CCSC,CNRS]{Yi-Fang Liu}
\ead{yifang.0731@gmail.com}
\author[CME,CCSC]{Wei Zhang}
\ead{weiz@tju.edu.cn}
\author[CME,CCSC]{Chao Xu \corref{cor}}
\ead{xuchao\_201310@126.com} %
\author[CNRS]{J{\o}rgen Vitting Andersen}
\ead{jorgen-vitting.andersen@univ-paris1.fr}
\author[CME,CCSC]{Hai-Chuan Xu}
\ead{highchance.xu@gmail.com}
\cortext[cor]{Corresponding author. Address: Rm.713 Beiyang Science Building, College of Management and Economics,
              Tianjin University, Tianjin 300072, China, Phone: +86 22 27891308.}

\address[CME]{College of Management and Economics, Tianjin University, Tianjin, 300072, China}
\address[CCSC]{China Center for Social Computing and Analytics, Tianjin University, Tianjin 300072, China}
\address[CNRS]{CNRS, Centre d'Economie de la Sorbonne, Universit\'{e} Paris 1 Panth\'{e}on-Sorbonne, 106-112 Boulevard de l'H\^{o}pital 75647 Paris Cedex 13 , France}

\begin{abstract}
  This paper studies the switching of trading strategies and its effect on the market volatility in a continuous double auction market. We describe the behavior when some uninformed agents, who we call switchers, decide whether or not to pay for information before they trade. By paying for the information they behave as informed traders. First we verify that our model is able to reproduce some of the stylized facts in real financial markets. Next we consider the relationship between switching and the market volatility under different structures of investors. We find that there exists a positive relationship between the market volatility and the percentage of switchers. We therefore conclude that the switchers are a destabilizing factor in the market. However, for a given fixed percentage of switchers, the proportion of switchers that decide to buy information at a given moment of time is negatively related to the current market volatility. In other words, if more agents pay for information to know the fundamental value at some time, the market volatility will be lower. This is because the market price is closer to the fundamental value due to information diffusion between switchers.
\end{abstract}


\end{frontmatter}

\section{Introduction}
\label{S1:Introduction}

This paper addresses the question of how \emph{pay-for-switch} behavior can affect financial markets in a continuous double auction mechanism. More precisely, we study how the percentage of switchers and the frequency of the switching are related to the volatility of the financial market. Franke and Westerhoff\cite{Franke-Westerhoff-2012-JEDC} have pointed out that there are indeed reasons that may result in the traders\textquoteright{} switching, such as herding, strategy fitness, i.e. pursuit of higher returns according to the past performance of the strategies and so on. We introduce a new switching mechanism that uninformed traders can become informed by paying for certain information cost. Although the impacts of the switching based on some fitness function or performance measure along the Brock and Hommes\textquoteright{} framework have been studied\cite{Brock-Hommes-1997-Econometrica,Brock-Hommes-1998-JEDC}, this paper is the first research considering \emph{pay-for-switch} behavior in a limit order market model. We try to find the relationship between the switching and the volatility of a limit order market. Since the introduction of the idea of excess volatility\cite{Shiller-1981-AER,LeRoy-Porter-1981-Eonometrica}, a large literature has been devoted to the topic. Behavioral finance tells us that the behavior of investors can affect the financial market, and switching is seen as one of the learning processes which could have some effect on the market volatility. Meanwhile, some scholars use switching to describe the learning behavior of investors in agent-based models and have studied its impact on the price dynamics. More recently, some researchers focus on agent-based models using heterogeneous information.

We initiate our research using an agent-based model in a continuous double auction market. The switching mechanism in our experiments is also based on heterogeneous information. There are four types of agents in our model: informed agents, uninformed agents, switchers and noise agents. We consider the pursuit of higher earnings according to the past performance as the reason for switching. At a given time a switcher considers whether he would have earned more if he paid for the information in the last time step. If so a switcher pays for the information and acts as an informed agent. Otherwise the switcher will not buy the information and acts as an uninformed agent. Next we study the stylized facts of our simulated market prices in order to test the validity of the model which are: fat tails of price returns, volatility clustering, no arbitrage (zero autocorrelation of returns) and long memory (slow decay of volatility autocorrelations)\cite{Gould-2012-OUWP}.

It has been proved that several parameters related to switching can affect the price dynamics of the market\cite{Brock-Hommes-1997-Econometrica}. However, whether switching can aggravate or reduce the market volatility is still unknown. We therefore pose the question: does the switching between different types of agents aggravate or reduce the market volatility, or exhibit different character under different market conditions? In our model, we consider the volatility of the market under different structures of agents. In other words, we validate whether the volatility is higher when there are more switchers in the market. Next we consider how, for a given fixed percentage $\rho$ of switchers, does the market volatility depend on proportion $\gamma$ of switchers \emph{actively} buying information at a given instant of time? Our general findings show that the larger the percentage $\rho$ of switchers the larger the volatility. However, the larger the percentage $\gamma$ of switchers paying for information at a given time, the lower will be the market volatility. This is different from the story in \cite{Brock-Hommes-1998-JEDC}. In our paper, this is because the switchers promote the diffusion of information and stabilize the market. Similar results have been obtained by \cite{di2013herding}, which finds that an initial increase in switching reduces the price volatility, but the effect becomes opposite when the switching increases further.

\section{Related literature}
\label{S1:RelatedLiterature}

Ever since the first ASMs used to study financial markets, a growing literature now describes how to study the price dynamics of markets caused by the investors\textquoteright{} behavior via agent-based models. In general, the behavior of the agents is time-varying and agents can choose different trading strategies according to some rules. Brock and Hommes\cite{Brock-Hommes-1997-Econometrica} create an agent-based model where traders can switch between different types and found that the dynamics of market prices are caused by a change in the intensity of choice to switch predictors. Since its introduction several papers have elaborated on the model and studied relevant questions. Chiarella and He\cite{Chiarella-He-2002-CE} conclude that the dynamics of pricing is affected by the relative risk attitudes of different types of investors (measured by the ratio of the relative risk aversion coefficients). Lux\cite{Lux-1995-EJ} designs a switching index that influences the probabilities with which the agents switch between different types. Lux and Marchesi\cite{Lux-Marches-2000-IJTAF} show that agents\textquoteright{} switching between fundamentalist and chartist strategies is the main reason that leads to volatility clustering and the emergence of fat-tailed returns. Similar results are found in other models\cite{Brock-Hommes-1998-JEDC,Youssefmir-Huberman-1997-JEBO}.

However, in general the agents in the afore-mentioned models are of the same two types: fundamentalists and chartists. The switching between different trading strategies can be considered for the following four reasons: predisposition as a behavioral bias, hypothetically differential wealth using the two different strategies over the past, herding, and a misalignment correctional mechanism\cite{Franke-Westerhoff-2012-JEDC}.

In our model we design a new type of agent called switchers who behave as follows: if the switchers paid for information last time and thereby earned more, they will pay for the information this time and therefore act as an informed trader. To the contrary they will not pay for the information and will therefore try to predict the prices as uninformed agents. The interpretation of the behavior mentioned above can be seen as a decision on whether or not to buy analysts\textquoteright{} reports in the real world. The viewpoint that research reports of securities and investment analysts can help the investors to get higher earnings has often been mentioned in the literature. For example, Dawson\cite{Dawson-1982-JPM} finds that one can get excess earnings if you follow the analysts\textquoteright{} report. Lee\cite{Lee-1987-JBFA} considers that the value of analysts\textquoteright{} reports is positively associated with the information in the reports. Further, Busse et al.\cite{Busse-Green-Jegadeesh-2010-JFM} use the empirical data to show that some investors who indeed follow the analysts\textquoteright{} reports thereby also change their behavior. That means some investors buy the reports and trade as the reports suggest.

Our model is based on heterogeneous information. Related literature\cite{Grossman-Stiglitz-1980-AER,Grundy-Kim-2002-JFQA,Kirchler-Huber-2007-JEDC} has shown that financial markets are indeed affected by given heterogeneous information that the agents possess. Grossman and Stiglitz\cite{Grossman-Stiglitz-1980-AER} prove that if costly information is immediately impounded into price, agents should not acquire it. However, the argument depends on how agents profit from their information, so the results are specific to a given price formation mechanism. Meanwhile, they consider whether agents who acquire information about the asset also affect the choice of trading strategies they employ. Grundy and Kim\cite{Grundy-Kim-2002-JFQA} find that the informational role of prices contributes positively to their variability, and the price variability is 20\% to 46\% higher than in an equivalent economy in which signals are publicly received in a heterogeneous information economy. Kirchler and Huber\cite{Kirchler-Huber-2007-JEDC} show that the heterogeneity of fundamental information is the driving force for trading, volatility, and ultimately the emergence of fat tails.

Once we have established a model, the validity of the model must be tested. Gould et al.\cite{Gould-2012-OUWP} point out that fat tails, autocorrelation and long memory can be main factors to test the validity of the model. Meanwhile, volatility clustering is also tested in many econometric and agent-based models\cite{Alfarano-Lux-Wagner-2005-CE,Lux-Schornstein-2005-JME,Weinbaum-2009-JEDC}. Therefore, we test the above four factors to check our model\textquoteright{}s validity.

Our model is used to validate the relationship between switching and market volatility, and much attention has been paid to the question that whether asset prices exhibit excess volatility in empirical studies. A large literature have proved that stock market prices have a higher volatility than the fundamental value of the market, and it is named excess volatility\cite{Shiller-1981-AER,LeRoy-Porter-1981-Eonometrica,Campbell-Cochrane-1999-JPE}. Bulkley and Tonks\cite{Bulkley-Tonks-1989-EJ} are the first authors to suggest learning as a possible explanation of excess volatility. Timmermann\cite{Timmermann-1996-RES} point out that the parameters of the dividend process are important for the dynamic behavior of stock prices, and conclude that the agents\textquoteright{} learning may generate predictability in stock returns and significantly increase the volatility of stock prices. Corrado et al.\cite{Corrado-Marcus-Lei-2007-IJFE} consider that asset mispricing may reflect investor psychology, and that excess volatility can arise from switching of sentiments. They show that excess volatility can be generated by the repeated entry and exit of currency \textquoteleft{}bulls\textquoteright{} and \textquoteleft{}bears\textquoteright{} with switches driven by \textquoteleft{}draw-down\textquoteright{} trading rules. Bullard and Duffy\cite{Bullard-Duffy-2001-MD} find that the observed excess volatility of asset returns can be explained by changes in investors\textquoteright{} expectations against a background of relatively small changes in fundamental factors. Wohlmuth and Vitting Andersen\cite{Wohlmuth-Vitting-Andersen-2006-PA} consider the competition of different traders acting on different time scales and with different information as yet another source of excess volatility. However none of such studies has considered the impact of switching whereas we suggest that the switching caused by switchers have a profound relationship with the excess volatility of the stock market prices. In addition we confirm the mechanism of switching has a different impact on the market under different market conditions.

\section{The Model}
\label{S1:Model}

In the following we will introduce a dynamic trading model with a single financial asset in a continuous double auction market. The market has a given tick size $\Delta$ as well as a transaction cost $\mu$ in the market. There are four types of agents: informed traders, uninformed traders, switchers and noise traders. All types of agents have their specific rules to trade and all the agents are risk neutral. All agents will entry the market at every time step. Every agent will decide how many orders he will submit in one time step. This is settled by a Poisson process with given parameter $\lambda$. For example, when a random number generated by the Poisson process is 2, the agent will submit two orders in this time step, and for each order he only trades one share stock; while when a random number generated by the Poisson process is 0, the agent will submit no orders in this time step. If the order is not executed, the agent will cancel his previous order when he re-entries to the market. Before participating in the market, switchers decide whether they want to become informed. Switchers then trade after having made their decision. First we validate the model with respect to the four stylized factors that we mentioned above. Then we use our model to consider the relationship between switching and the market volatility. Further details about the elements of the model are as follows:

\subsection{Asset Pricing}
\label{S2:AssetPricing}

There is only one asset in the market, and the asset has a fundamental value $v_t$ at a given instant time $t$. The fundamental value is the expectation value of the present asset and evolves according to a Poisson process $N(t)$ with parameter $\varphi$,

\begin{equation}\label{Eq:CommonValue}
    v_{t+1} = v_t + \sum^{N(t)} \delta,
\end{equation}
where $\delta\in(-\Delta, \Delta)$ is selected with equal probability.

The market price $p_t$ is the average trading price at time $t$, as follow,

\begin{equation}
 p_t = \frac{\sum_n p_n^t \times s_n}{\sum_n s_n},
 \label{Eq:MarketPrice}
\end{equation}
$p_n^t$ means the price of each executed order $n$ at time $t$, and $s_n$ means the order size (one unit, two units, or more) for each executed order $p_n^t$ at time $t$. If there is no trading at time $t$, $p_t=p_{t-1}$, and the initial market price is $v_0$.

\subsection{Heterogeneous Traders}
\label{S2:HeterogeneousTraders}

There are four types of agents in the market including informed agents, uninformed agents, switchers and the noise traders. However, the different agents have different opinions on the prediction of the asset price. The informed traders know the current fundamental value of the asset, and they predict the fundamental value as the market price of the asset. The uninformed agents know the fundamental value with a time lag, and they use Genetic Algorithms (GA) to predict the market price. The switchers a priori act as uninformed agents but can switch to an informed agent if they judge the cost to do so is low enough (for details see below). If they pay for the information, then they know the current fundamental value and switch and trade as the informed agents. To the contrary, they use past prices to predict the current asset price and trade as uninformed agents. The noise agents predict the price as the uninformed agents, but they can\textquoteright{}t learn from past price behavior of the market and their coefficients of the formulae for predicting the market price are random. Specifically the different agents trade accordingly:

\begin{enumerate}
  \item Informed traders

  The informed agents in our model represent fundamentalist traders in a real market. This kind of agent knows the current fundamental value $v_t$, and they predict the asset price $p_{I,t}^e$ as $v_t$.

  \item Uninformed traders

  The uninformed agents in our model represent traders who predict the stock price by using their own technology and information. The uninformed agents view $v_t$ with a time lag $\tau$, that is, an uninformed agent in the market at time $t$ just knows $v_{t-\tau}$. The symbol $p_¦Ó^{ave}$ denote the average transaction price over the past $\tau$ time steps and $p_m$ means the midpoint of the bid-ask quote at the current time.

  Uninformed agents predict the asset price using Equation ~\ref{Eq:PredictPrice}.

  \begin{equation}
    p_{U,t}^e = \frac{a^i \cdot v_{t-\tau} + b^i \cdot p_t^{ave} + c^i \cdot p_m}{a^i + b^i + c^i },
    \label{Eq:PredictPrice}
  \end{equation}
  where $a^i$, $b^i$ and $c^i$ are the coefficients in the range [0,1]. The uninformed agents use Genetic Algorithms to optimize the three coefficients continuously according to the market conditions.

  \item Zero-intelligence traders

  The zero-intelligence agents represent some blind market participants. They also utilize Equation ~\ref{Eq:PredictPrice} to predict asset price but the coefficients are random.

  \item Switchers

  The switchers can change their behavior and act either as informed agents or uninformed agents. Since the information of knowing the fundamental value $v_t$ is valuable they decide whether or not to purchase that information. They do so by comparing the error made in the last time step predicting the market price $p_{t-1}$ using $v_{t-1}$ , to the error made predicting $p_{t-1}$ using $p_{U,t-1}^e$. If the information cost plus the error forecasting $p_{t-1}$ using $v_{t-1}$ is larger than the error forecasting $p_{t-1}$ using $p_{U,t-1}^e$ , the agent will stick to his forecast using $p_{U,t}^e$ . Otherwise the agent will buy the information and use $v_t$ to forecast. To summarize:

  \begin{align}\label{Eq:SwitchPrediction}
  \vspace{-5mm}
  p_{S,t}^e = \begin{cases}
   (a^i \cdot v_{t-\tau} + b^i \cdot p_t^{ave} + c^i \cdot p_m)/(a^i + b^i + c^i ) &\mbox{if~} |p_{U,t-1}^e-p_{t-1}| < |v_{t-1}-p_{t-1}| + C\\
    v_t & \mbox{if~} |p_{U,t-1}^e-p_{t-1}| \geq |v_{t-1}-p_{t-1}| + C
  \end{cases}
  \end{align}
  $C$ is the information cost. All the other symbols are the same as Equation ~\ref{Eq:PredictPrice}. If the agent didn't buy information at time $t-1$, which means he don't know $v_{t-1}$ at time $t$, then he simply treats $p_{t-1}$ as $v_{t-1}$, so that he can also use Equation ~\ref{Eq:SwitchPrediction} to make his switching decision. Note that, when the switcher switches from informed to be uninformed at time $t$, he will ignore $v_{t-1}$ and still use $v_{t- \tau}$ to form his forecast, because no evidence shows that using $v_{t-1}$ instead of $v_{t- \tau}$ can obtain more accurate $p_U^e$.
\end{enumerate}

\subsection{Order Submission Rules}
\label{S2:OrderSubmission}

Traders' order submissions are based on rules constructed by \cite{gil2007price,wei2013learning}, where traders who submit limit orders need a liquidity compensation, which is at least able to cover the transaction cost $\mu$. In other words, buy limit order price can be obtained by subtracting $\mu$ from $p_t^e$; sell limit order price can be obtained by adding $\mu$ on $p_t^e$. The order submission rules are listed in Table~\ref{Tab:OrderSubRules}.

\begin{table}[!htb]
\centering
\begin{threeparttable}
  \caption{\label{Tab:OrderSubRules} Order submission rules.}
  \begin{tabular}{l l}
  \toprule
    Scenario    & Order \\ \midrule
    \multicolumn{2}{l}{\emph{Case 1: There is at least one ask price and one bid price in the limited order book}}                                      \\
    ${p_t^e > ask + \mu}$                                                                           & Market order to buy                               \\
    ${ask + \mu \geq p_t^e \geq bid - \mu \& \mid ask - p_t^e \mid \leq \mid p_t^e - bid \mid}$     & Limit order to buy with ${p_l = p_t^e - \mu}$     \\
    ${ask + \mu \geq p_t^e \geq bid - \mu \& \mid ask - p_t^e \mid > \mid p_t^e - bid \mid}$        & Limit order to sell with ${p_l = p_t^e + \mu}$    \\
    ${p_t^e < bid - \mu}$                                                                           & Market order to sell                              \\
    \multicolumn{2}{l}{\emph{Case 2: There are no bid prices}}                                                                                          \\
    ${p_t^e > ask + \mu}$                                                                           & Market order to buy                               \\
    ${p_t^e \leq ask + \mu}$                                                                        & Limit order to buy with ${p_l = p_t^e - \mu}$     \\
    \multicolumn{2}{l}{\emph{Case 3: There are no ask prices}}                                                                                          \\
    ${p_t^e < bid - \mu}$                                                                           & Market order to sell                              \\
    ${p_t^e \geq bid - \mu}$                                                                        & Limit order to sell with ${p_l = p_t^e + \mu}$    \\
    \multicolumn{2}{l}{\emph{Case 4: There are no ask or bid prices}}                                                                                   \\
    With probability 50\%                                                                           & Limit order to buy with ${p_l = p_t^e - \mu}$     \\
    With probability 50\%                                                                           & Limit order to sell with ${p_l = p_t^e + \mu}$    \\
  \bottomrule
  \end{tabular}
  \begin{tablenotes}
    \footnotesize
    \item $ask$ denotes the best ask price; $bid$ denotes the best bid price; $p_l$ denotes limit order price.
  \end{tablenotes}
\end{threeparttable}
\end{table}

\subsection{Switching Mechanism}
\label{S2:SwitchingMechanism}

Since the asset price by definition fluctuates around the fundamental value at all times, the average earnings of informed agents will be higher than the uninformed ones. As described before, the uninformed agents hope they can get higher profits by buying the information about $v_t$. However in order to decide whether an agent will switch or not we first need to calculate the cost of information $C$.

\begin{figure}[!htb]
  \centering
  \includegraphics[width=12cm]{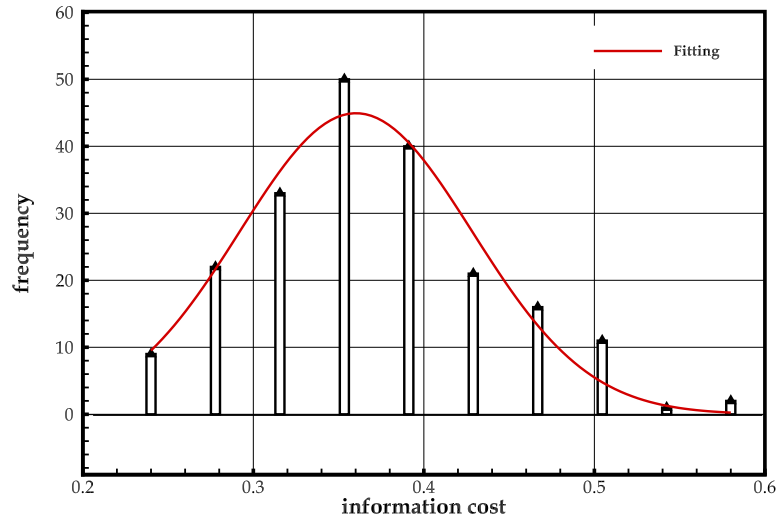}
  \caption{\label{Fig:Info:Cost} Distribution of information cost $C$ obtained from simulating $200$ artificial markets without switchers. The information cost data conform to the normal distribution with function $f(x)=a \cdot \exp(-(\frac{x-b}{c})^2)$. $a=44.83$, $b=0.3604$, $c=0.09662$, and the adjust $R^2$ is $0.9362$. }
\end{figure}

\begin{figure}[!htb]
  \centering
  \includegraphics[width=12cm]{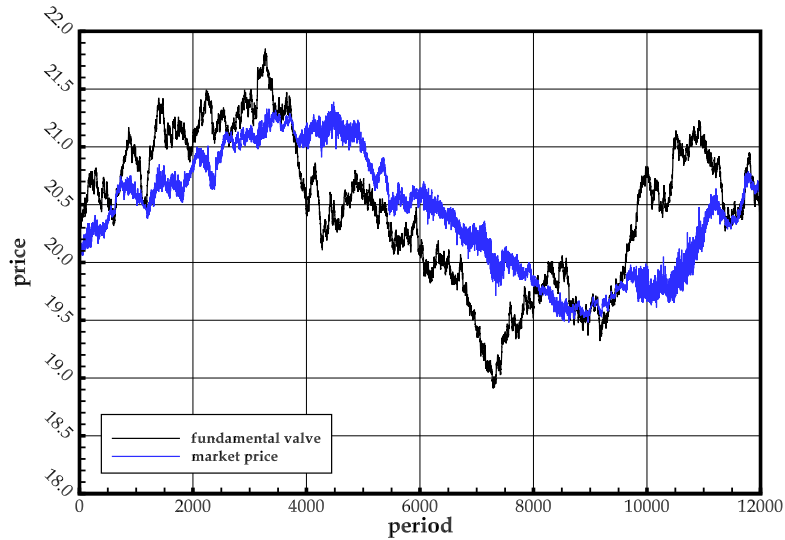}
  \caption{\label{Fig:Fun:Mar:Price} Plot of the fundamental value and the market price. The black line represents the fundamental value while the blue one refers to the market price.}
\end{figure}

We solve this problem as follows: We first make some runs of the artificial market but without the presence of switchers. The parameter values of the artificial market are listed in Table~\ref{Tab:Parameter}. After each run we define the difference between the informed agents\textquoteright{} order profits and the uninformed agents\textquoteright{} order profits as the information cost $C$. The order profits are defined as

\begin{equation}\label{Eq:OrderProfits}
    h =
    \left\{
    \begin{array}{rl}
    p_l-v_t, & for\ sell\ limit\ order\ \\
    v_t-p_l, & for\ buy\ limit\ order\ \\
    p_t-v_t, & for\ sell\ market\ order\ \\
    v_t-p_t, & for\ buy\ market\ order\
    \end{array}
    \right.
\end{equation}
We run the artificial market for $200$ times to design the difference as a distribution of $C$.

\begin{table}[!ht]
  \caption{\label{Tab:Parameter} Parameter values.}
  \centering
  \begin{tabular}{c c}
  \toprule
  Parameters    & Values    \\ \midrule
  $v_0$         & 20        \\
  $\Delta$      & 0.01  \\
  $\mu$         & 0.04  \\
  $\varphi$     & 4 \\
  $\lambda$     & 1  \\
  $\tau$        & 1200 \\
  \bottomrule
  \end{tabular}
\end{table}

The simulation result of distribution for $C$ is shown in figure~\ref{Fig:Info:Cost}. Apparently, it looks like a normal distribution. Therefore we use normal distribution to fit the distribution of $C$. $R^2$ is about $0.95$ and the adjusted $R^2$ is $0.9362$, so we can conclude that the distribution is indeed a normal distribution. The average information cost is 0.36 and has an extremely small probability to be negative, which is consistent with the analysis above. The estimated information cost will be used for following simulations.

\subsection{Testing our model}
\label{S2:TestingModel}

As mentioned in the introduction, empirical research has found that switchers indeed buy information and then follow the recommendations. We have designed a model to describe the behavior. However, whether the model gives a reliable description of real market data should first be validated. We test the model for the following four factors: fat tails, autocorrelation, long memory and volatility clustering.

\begin{figure}[!htb]
  \centering
  \includegraphics[width=12cm]{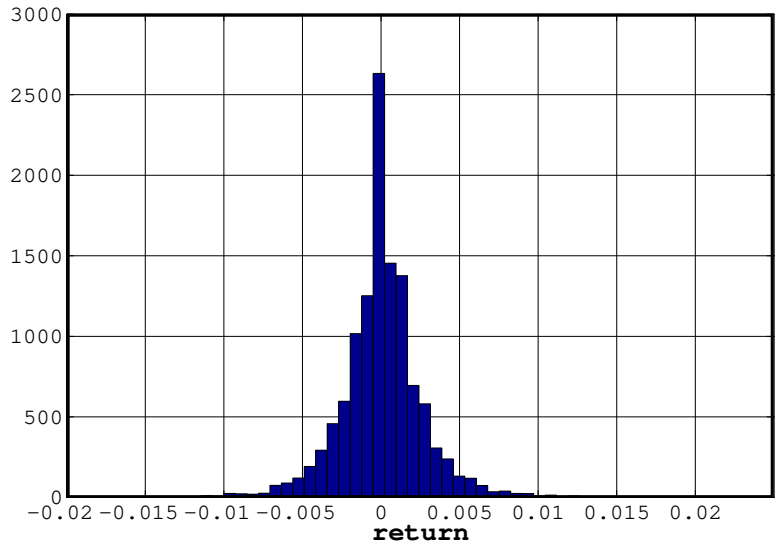}
  \caption{\label{Fig:Ret:Hist} Histogram and of the return of the market price. The return of the market price is calculated by each time step using the data in Figure~\ref{Fig:Fun:Mar:Price}. Leptokurtosis is obvious in this plot.}
\end{figure}

\begin{table}[!ht]
  \caption{\label{Tab:DesStatProperty} Descriptive statistical property.}
  \centering
  \begin{tabular}{c c}
  \toprule
  \multicolumn{2}{c}{Descriptive Statistical Property} \\ \midrule
  Mean          & -1.94e-06 \\
  Median        & 0.000000  \\
  Maximum       & 0.006246  \\
  Minimum       & -0.006282 \\
  Std. Dev.     & 0.001038  \\
  Skewness      & -0.050051 \\
  Kurtosis      & 6.632679  \\
  Jarque-Bera   & 6602.640  \\
  Probability   & 0.000000  \\
  \bottomrule
  \end{tabular}
\end{table}

Figure~\ref{Fig:Fun:Mar:Price} shows a plot of the fundamental value and the market price over 12000 time steps. Figure~\ref{Fig:Ret:Hist} shows the histogram of the returns of market price. From figure~\ref{Fig:Fun:Mar:Price} we can see that the market price (blue line) fluctuates closely around the behavior of the fundamental value (black line).

In Table~\ref{Tab:DesStatProperty}, we show some statistics characteristics of the returns of the market price, from which one can see the fat-tailed character of the returns quite clearly.

\begin{figure}[!htb]
  \centering
  \includegraphics[width=12cm]{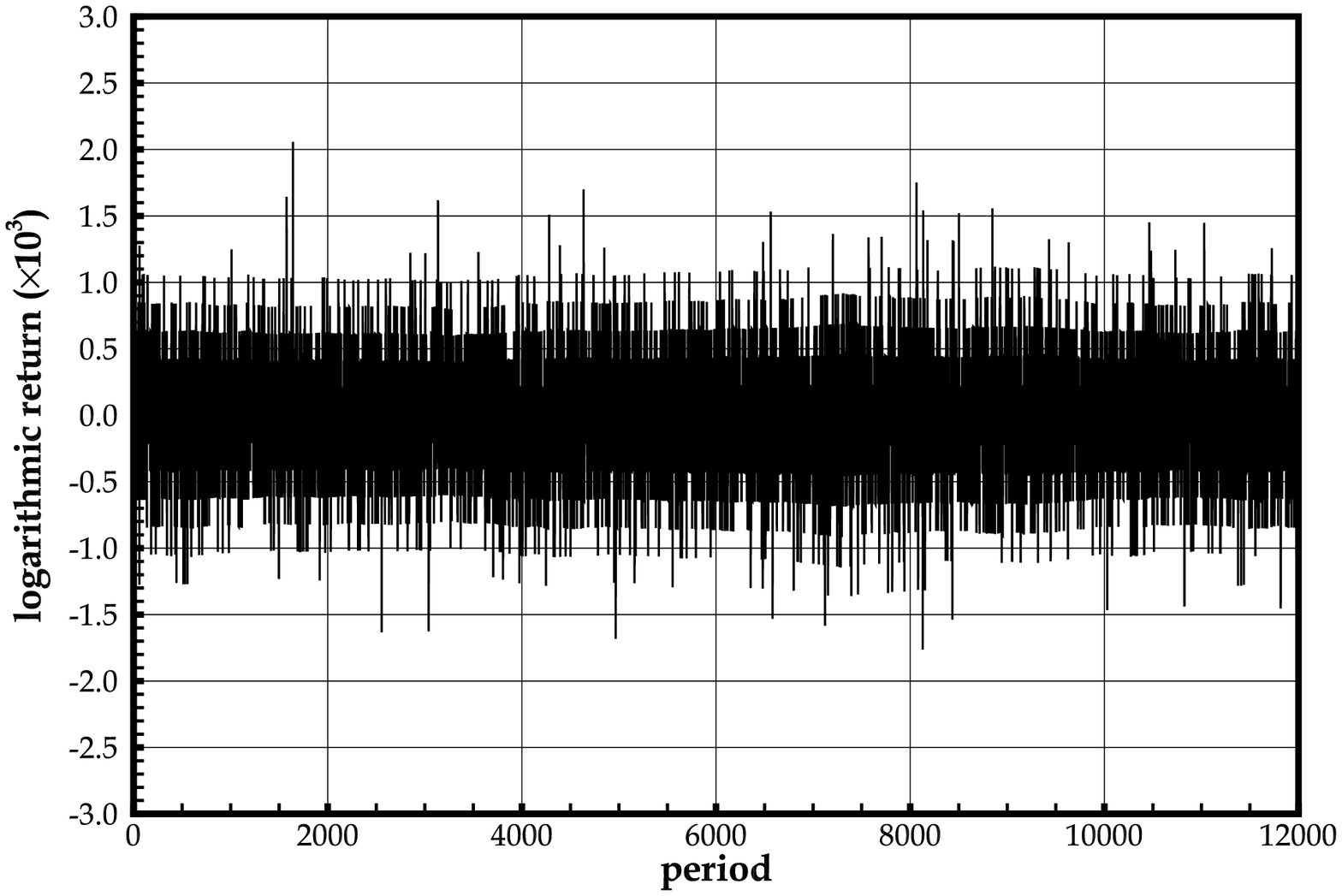}
  \caption{\label{Fig:Log:Ret:Fun} Example of a time series showing the logarithmic returns of fundamental value. The logarithmic return is between -0.002 and 0.002 all the time and there is no volatility clustering. The simulation illustrates a white-noise process.}
\end{figure}

\begin{figure}[!htb]
  \centering
  \includegraphics[width=12cm]{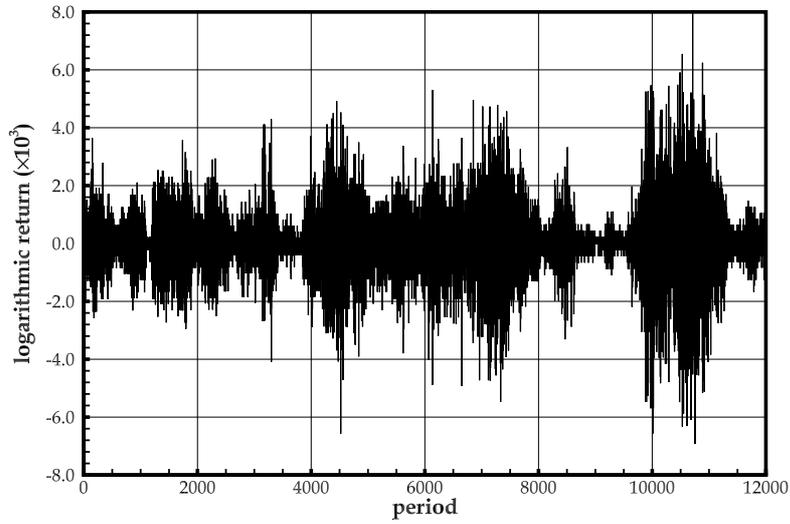}
  \caption{\label{Fig:Log:Ret:Mar} Example of a time series showing the logarithmic returns of the market price Compare to Figure~\ref{Fig:Log:Ret:Fun}, the amplitude of the logarithmic return of market price is larger than the one of fundamental value. Meanwhile, volatility clustering is obvious in this figure.}
\end{figure}

\begin{table}[!ht]
  \caption{\label{Tab:ArchTestFun} ARCH test for fundamental value.}
  \centering
  \begin{tabular}{c c c c}
  \toprule
  \multicolumn{4}{c}{ARCH test for fundamental value} \\ \midrule
  F-statistic   & 1.262826  & Probability   & 0.261139  \\
  Obs*R-squared & 1.262904  & Probability   & 0.261102  \\
  \bottomrule
  \end{tabular}
\end{table}

\begin{table}[!ht]
  \caption{\label{Tab:ArchTestCoeFun} The coefficients of ARCH test equation for fundamental value.}
  \centering
  \begin{tabular}{c c c c c}
  \toprule
  Variable      & Coefficient   & Std. Eror     & t-Statistic   & Probability   \\ \midrule
  a             & 2.25e-07      & 3.80e-09      & 59.14381      & 0.000000      \\
  b             & 0.010260      & 0.009130      & 1.123755      & 0.261100      \\
  \bottomrule
  \end{tabular}
\end{table}

\begin{table}[!ht]
  \caption{\label{Tab:ArchTestMar} ARCH test for market price.}
  \centering
  \begin{tabular}{c c c c}
  \toprule
  \multicolumn{4}{c}{ARCH test for market price} \\ \midrule
  F-statistic   & 512.8676  & Probability   & 0.000000  \\
  Obs*R-squared & 0.5771    & Probability   & 0.000000  \\
  \bottomrule
  \end{tabular}
\end{table}

\begin{table}[!ht]
  \caption{\label{Tab:ArchTestCoeMar} The coefficients of ARCH test equation for market price.}
  \centering
  \begin{tabular}{c c c c c}
  \toprule
  Variable      & Coefficient   & Std. Eror     & t-Statistic   & Probability   \\ \midrule
  a             & 9.21e-07      & 2.54e-09      & 36.20055      & 0.000000      \\
  b             & 0.202494      & 0.008941      & 22.64658      & 0.000000      \\
  \bottomrule
  \end{tabular}
\end{table}

Figure~\ref{Fig:Log:Ret:Fun} shows as a function of time the logarithmic return of fundamental value, whereas figure~\ref{Fig:Log:Ret:Mar} shows the market price. Volatility clustering is clearly present in figure~\ref{Fig:Log:Ret:Mar}, so volatility clustering is indeed generated by the agents\textquoteright{} trading just as observed in real markets. In addition we use ARCH-LM to test for volatility clustering in our agent-based model. The ARCH-LM test equation is as

\begin{equation}\label{Eq:ARCHTest}
    \varepsilon_t^2 = a + b * \varepsilon_{t-1}^2
\end{equation}
where $\varepsilon$ denotes the residuals estimated by AR(1) model. The results are shown in Table~\ref{Tab:ArchTestFun}, Table~\ref{Tab:ArchTestCoeFun}, Table~\ref{Tab:ArchTestMar} and Table~\ref{Tab:ArchTestCoeMar}. Table~\ref{Tab:ArchTestFun} and Table~\ref{Tab:ArchTestCoeFun} shows that there are no ARCH effects of the logarithmic return of the fundamental value while from Table~\ref{Tab:ArchTestMar} and Table~\ref{Tab:ArchTestCoeMar} we can see that there are ARCH effects of the logarithmic returns of the market price.


\begin{figure}[!htb]
  \centering
  \includegraphics[width=12cm]{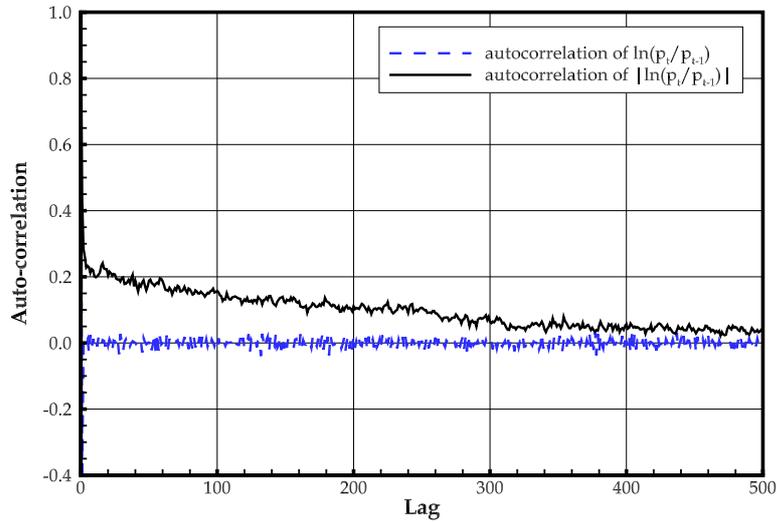}
  \caption{\label{Fig:Autocorr:Mar} Auto-correlations of the returns and absolute returns of the market price. The black line refers to the auto-correlations of the absolute returns and the blue line means the auto-correlations of returns. There is no auto-correlations between the returns of market price, while the absolute returns of the market price show slow decaying auto-correlations.}
\end{figure}

Next, we consider the auto-correlations of the returns of the market price. Figure~\ref{Fig:Autocorr:Mar} shows the auto-correlations of the returns and absolute returns of the market price. We can see clearly that there is no auto-correlations in the returns of the market price (no arbitrage possibility). However, we find that the absolute returns of the market price indeed show slow decaying auto-correlations. Then we conclude that the characteristic of the returns is similar to the real market.

The GARCH model gives a way to test the short-term memory of the logarithmic price returns, while the Hurst exponent can be used to test the long-term memory. We run the artificial market for 5 times to test the Hurst exponent of the logarithmic price returns of the fundamental value and the market price. More details are shown in Table~\ref{Tab:HustExpFunMar}. We find that the Hurst exponents of the returns of the fundamental value in the five experiments are about 0.5, while the Hurst exponents of the returns of the market price in the five experiments are far higher more than 0.5. Therefore we can conclude that there is no long-term memory in the series of logarithmic fundamental value returns, but the series of logarithmic market returns has long-term memory.

\begin{table}[ht]
  \caption{\label{Tab:HustExpFunMar} Hurst exponent of fundamental value and market price.}
  \centering
  \begin{tabular}{c c c}
  \toprule
                    & Fundamental Value & Market Price  \\ \midrule
    Simulation 1    & 0.5527            & 0.8435        \\
    Simulation 2    & 0.5771            & 0.8774        \\
    Simulation 3    & 0.5309            & 0.9079        \\
    Simulation 4    & 0.5576            & 0.8681        \\
    Simulation 5    & 0.5653            & 0.8003        \\
  \bottomrule
  \end{tabular}
\end{table}

\begin{table}[ht]
  \caption{\label{Tab:DiffPerAgents} Different percentages of agents.}
  \centering
  \begin{tabular}{c c c c c}
  \toprule
                    & Informed Traders  & Uninformed Traders    & Zero-Intelligence Traders & Switchers $(\rho)$    \\ \midrule
    Simulation 1    & 12\%              & 30\%                  & 58\%                      & 0\%                   \\
    Simulation 2    & 12\%              & 23\%                  & 58\%                      & 7\%                   \\
    Simulation 3    & 12\%              & 15\%                  & 58\%                      & 15\%                  \\
    Simulation 4    & 12\%              & 8\%                   & 58\%                      & 22\%                  \\
    Simulation 5    & 12\%              & 0\%                   & 58\%                      & 30\%                  \\
  \bottomrule
  \end{tabular}
\end{table}

\subsection{Computational design}
\label{S2:ComputationalDesign}

After testing our model, we consider the relationship between the switching and the dynamics of the market price. We design the experiments as follows:

We examine whether there is some relationship between the switching of strategies of the agents and the market volatility. To do so we design experiments with different configurations of switchers. Similar to \cite{wei2013learning}, we will fix the percentages of the informed and zero-intelligence agents to be 12\% and 58\%. In our model the uninformed agents are divided into switchers and uninformed traders who can not switch. We choose 5 different proportions evenly from 0\% to 30\% as switchers' percentages, and thus we consider 5 different model configurations as described in Table~\ref{Tab:DiffPerAgents}.

The idea is now to compare the volatility of the market price under different structures of agents. For each configuration mentioned in Table~\ref{Tab:DiffPerAgents}, we choose 30 random realizations of the market by fixing a random initial seed for each run of the market. For each random seed, we run the market under different structures of agents. Therefore, we can ensure to have the same different market conditions under different percentages of the switchers. If the results are different, we can conclude it is due to the different percentages and not different random seeds used in the simulations.

\section{Results}
\label{S1:Results}

\begin{figure}[!htb]
  \centering
  \includegraphics[width=8cm]{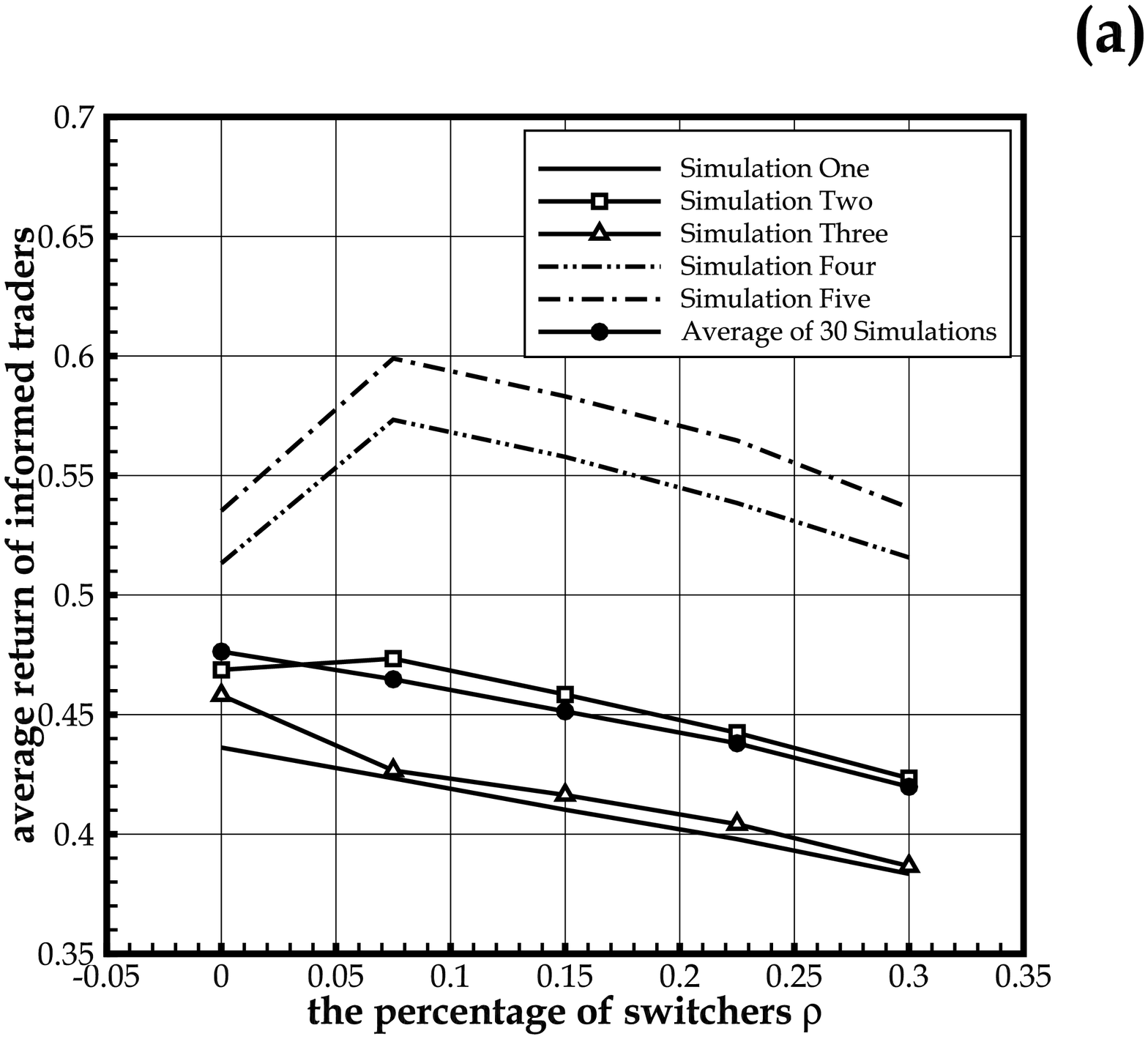}
  \includegraphics[width=8cm]{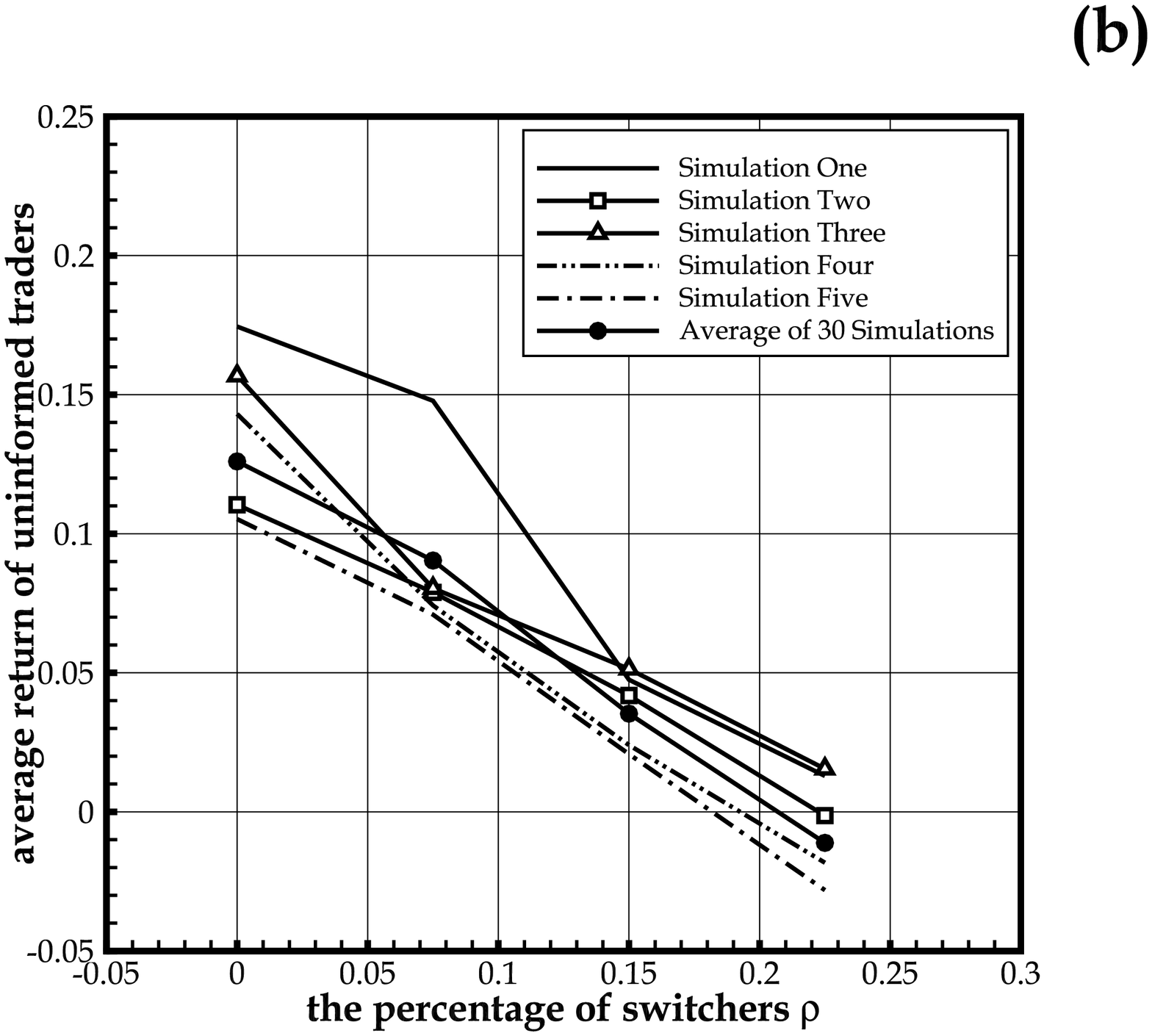}
  \includegraphics[width=8cm]{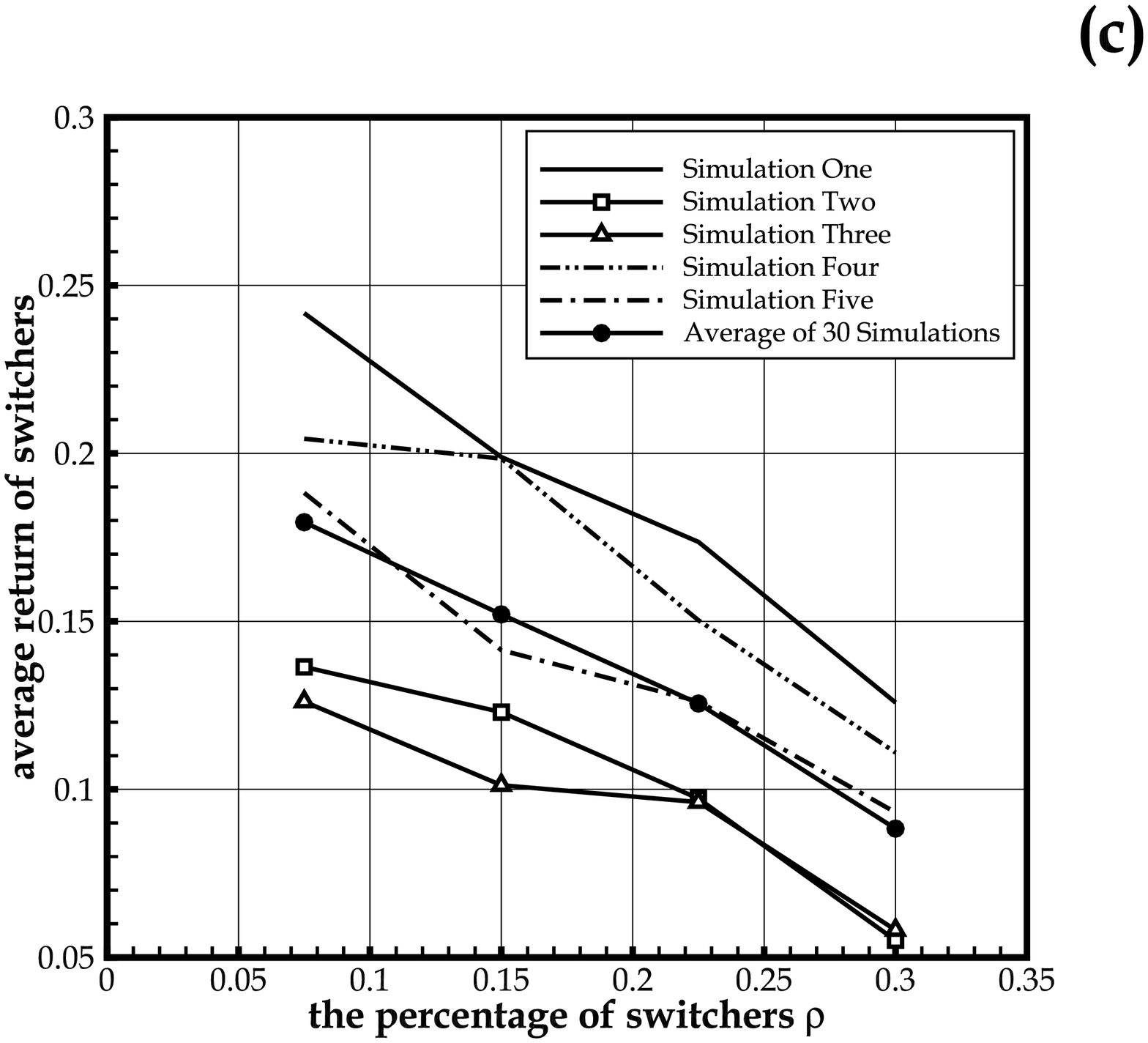}
  \caption{\label{Fig:Tra:Ret} Returns of Different Types of Agents versus $\rho$: (a) Informed Agents, (b) Uninformed Agents, and (c) Switchers. Thirty simulations are used to calculate the returns for each type of agent and five simulations are randomly chosen as shown in the figure. The same random seed is used in each simulation for the different types of agents. The returns of the informed traders, Plot (a), show a small declining tendency versus $\rho$. However, the return of the uninformed traders and switchers, Plot (b) and (c), show a much steeper decline versus the percentage of switchers $\rho$.}
\end{figure}

Figure~\ref{Fig:Tra:Ret}  shows the average order returns of different agents with different percentages of switchers $(\rho)$ corresponding to the 5 configurations of traders reported in Table~\ref{Tab:DiffPerAgents}. Each plot illustrates one type of agents and shows 5 typical market runs as well as the average of in total 30 market runs. For all kinds of traders, the average returns are reported before the transaction cost. But for switchers, the average returns reported in Figure~\ref{Fig:Tra:Ret}(c) are after the information cost so that the returns of different kinds of traders can be comparable.

We can see that there is almost no change in the informed agents\textquoteright{} return as a function of $\rho$ except a little decline (Figure~\ref{Fig:Tra:Ret}(a)). However in the case of uninformed agents the return declines more obviously versus $\rho$. There is no difference between the returns of informed agents while the returns in the other two types are obviously different under different structures of agents. Overall, switchers can choose to buy information under the condition that the improved predictive accuracy can compensate information cost, so that they earn more than uninformed agents.

\begin{figure}[!htb]
  \centering
  \includegraphics[width=12cm]{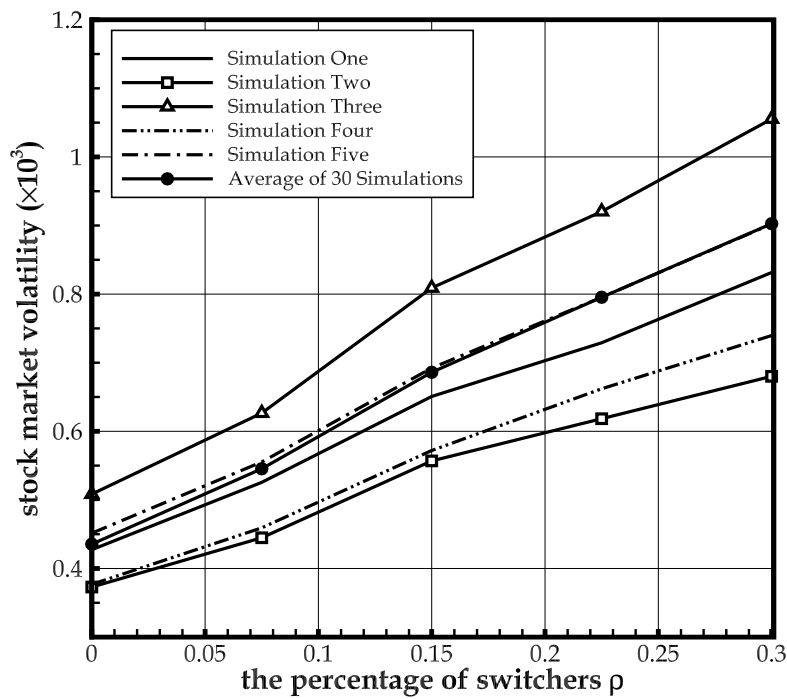}
  \caption{\label{Fig:Mar:Vol:Rho} Market volatility under different structures of agents. The stock market volatility shows a clear increase versus $\rho$.}
\end{figure}

\begin{figure}[!htb]
  \centering
  \includegraphics[width=12cm]{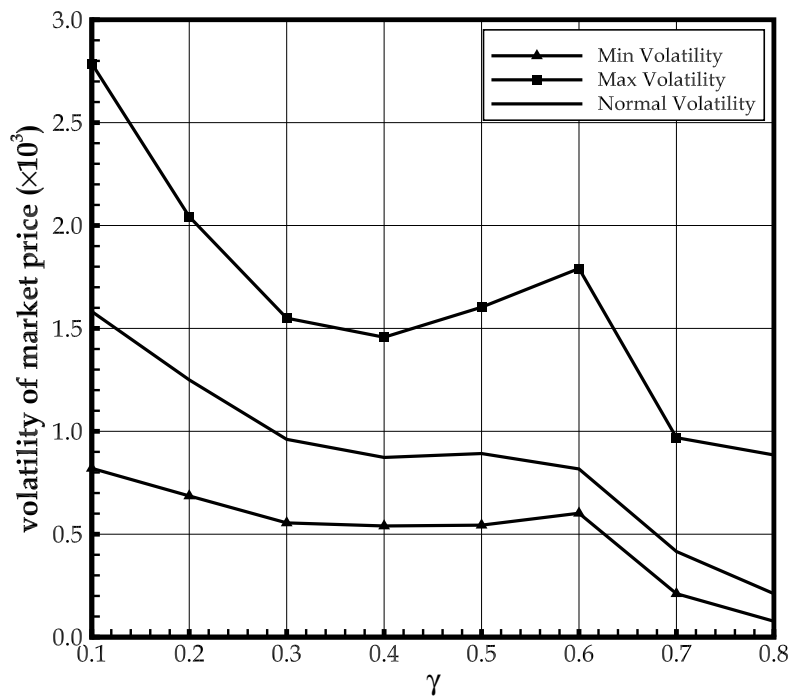}
  \caption{\label{Fig:Mar:Vol:Gama} Average Volatility under Different Proportions $\gamma$. $\gamma$ represents the proportion of switchers buying information at a given instant of time when the percentage $\rho$ of switchers is fixed ($\rho=30\%$). The volatility of the market price is a decreasing function of $\gamma$.}
\end{figure}

\begin{table}[ht]
  \caption{\label{Tab:AveMarVol} Average market volatility under different percentage of switchers.}
  \centering
  \begin{tabular}{c c c c c c}
  \toprule
  percentage of switchers $\rho$    & 0\%       & 7\%       & 15\%      & 22\%      & 30\%      \\ \midrule
  Average market volatility         & 0.00043   & 0.00054   & 0.00069   & 0.00079   & 0.00090   \\
  \bottomrule
  \end{tabular}
\end{table}

Figure~\ref{Fig:Mar:Vol:Rho} shows different market volatilities under different percentages of switchers. One notes that in each simulation the market volatility is the smallest in the case without switchers, and we can get the same conclusion when making the average over 30 simulations. The first general phenomenon we have shown is therefore that the market with switchers fluctuates more than the market with no such agents. The second one is that the larger the percentage of uninformed agents acting as switchers, $\rho$, the larger is the volatility of the market returns, see in Table~\ref{Tab:AveMarVol}. Above all, we can conclude that the behavior of switchers indeed has some effect on the market and they are a destabilizing factor in the market. We can explain such phenomena as follows: We find that the returns of the switchers are less with larger the percentage of switchers, which means that their forecasts deviate from the fundamental value more and become more inaccurate. Therefore the market volatility is higher and they become the destabilizing factor in the market.

We then consider for a given fixed percentage $\rho$ of switchers, how does the market volatility depend on proportion $\gamma$ of switchers \emph{actively} buying information at a given instant of time. Figure~\ref{Fig:Mar:Vol:Gama} shows the average market volatilities as a function of $\gamma$. The data is obtained averaging over 30 simulations using a fixed percentage of switchers at 30\%. We got similar results from the other experiments (see Table~\ref{Tab:DiffPerAgents}) with different fixed values of $\rho$. We can see that the volatility (minimum, maximum or normal) declines as the proportion of switchers \emph{actively} buying, $\gamma$, increases. Specifically, the market volatility is high when there are relatively few information buyers at a given instant of time ($\gamma$ below 20\%). On the contrary the market volatility is relatively low when $\gamma$ is above 70\%. In general we find that when most of the switchers buy information at a given time, the market volatility is low. Inversely, the market volatility is high if only a small percentage of agents buy. Therefore we conclude that the behavior of switching can affect the market, and if the \emph{active} switching rate is higher (meaning that more switchers become informed agents), the volatility of the market will be lower. This is because when more switchers switch at a given time, the diffusion of information is higher and the market price is close to the fundamental value. Therefore the market is more effective and thereby more stable.

\section{Conclusion}
\label{S1:Conclusion}

In this paper we have studied the reason why some uninformed agents in an agent based model switch to other types of agents, and analyzed the influence of switching on the market stability. As is shown in Busse et al.\cite{Busse-Green-Jegadeesh-2010-JFM}, some investors buy the analysts\textquoteright{} reports and follow their recommendations. We have designed a new type of agent based model where agents decide whether to pay for information to become informed before they make a trading decision. First we have validated our model and found characteristics (stylized facts) which are similar to the price dynamics observed in real markets. Secondly we have analyzed the influence of switching on the returns of the agents as well as on the market volatilities under different conditions.

The first conclusion is that the market volatility is larger if there are switchers in the market. We find that the market volatility without switchers is smaller than the market volatility with whatever the percentage of switchers. We therefore deduce that in general the behavior of agents switching between two types of trading strategies can increase the market volatility.

The second conclusion is that we find that the market volatility is higher the larger the percentage of switchers in the market. In addition, we find the returns of uninformed agents and switchers decline when there are more switchers, while there is almost no change concerning the informed agents\textquoteright{} return. So we conclude that switchers\textquoteright{} forecasts become more inaccurate the higher the percentage of switchers in the market, $\rho$, because their returns decrease with $\rho$. Therefore they become a destabilizing factor in the market due to the inaccurate forecasts. As a result the market volatility is higher with more switchers in the market.

The last conclusion is that the market volatility is lower if, for a fixed percentage of switchers in the market, there are more switchers paying actively for information in order to know the fundamental value at a given time. In other words, the percentage of switching is negatively correlated with the current market volatility.

We can explain this by the fact that the more switchers who know the fundamental value, the more will the market price be closer to the fundamental value. As a result, the market volatility is lower.
Overall, we find that the market volatility indeed increases with larger percentage of switchers in the market, but the market volatility will decrease with more switchers actively pay for information to know the fundamental value. Therefore, whether the behavior of switchers can increase or decrease the market volatility depends on the situation. The market volatility is lower if the switchers always pay for information to know the fundamental value, while the market volatility is higher if the switchers always switch between the two types of decision making. However, there are still unresolved issues concerning switching. We can just conclude that the pursuing of more earnings according to the past performance of the strategies is one of the reasons why traders switch. There are also other reasons we haven\textquoteright{}t considered in the article that may cause switches such as herding and so on. Therefore more research will indeed be needed to confirm the reason why traders switch between different types of decision making.

\bigskip
{\textbf{Acknowledgments:}}

We thank Editor H.E. Stanley and anonymous referees for useful suggestions that helped us improve the paper substantially. We also would like to express our heartfelt gratitude to Li-jian Wei, who provided useful comments to improve this paper. This work was supported by the National Natural Science Foundation of China under Key Project Grant No 71131007 and the Programme of Innovative Research Team supported by Ministry of Education of China under Grant No IRT1028.

\bibliography{infocost}

\end{document}